\documentclass[a4paper,12pt]{article}
\usepackage{moreverb}
\usepackage{natbib}
\usepackage{graphicx,amssymb,amstext,amsmath}
\usepackage{float}
\usepackage{caption}

\usepackage[caption = false]{subfig}
\newcommand\ie{\textsl{i.e.}\ }

\def\s{\sigma^2}

\def\m{\mu}
\def\m1{\mu_1}
\def\m0{\mu_0}

\def\ci{_{1i}}

\def\Yc{Y_{1i}}
\def\Ye{Y_{2i}}
\def\ci{\perp\!\!\!\perp}

\begin{document}

\title{Non-compliance and missing data  in health economic evaluation}
\author{Karla ~DiazOrdaz,  Richard ~Grieve\\
London School of Hygiene and Tropical Medicine}

\date{} 
\maketitle

\begin{abstract}
Health economic evaluations face the issues of non-compliance and missing data. Here, non-compliance is defined as non-adherence to a specific treatment, and occurs within randomised controlled trials (RCTs) when participants depart from their random assignment. Missing data arises if, for example, there is loss to follow-up, survey non-response, or the information available from routine data sources is incomplete. Appropriate statistical methods for handling non-compliance and missing data have been developed, but they have rarely been applied in health economics studies.  Here, we illustrate the issues and outline some of the appropriate methods to handle these with an application to a health economic evaluation that uses data from an RCT.

In an RCT the random assignment can be used as an instrument for treatment receipt, to obtain consistent estimates of the complier average causal effect, provided the underlying assumptions are met. Instrumental variable methods can accommodate essential features of the health economic context such as the correlation between individuals' costs and outcomes in cost-effectiveness studies. Methodological guidance for handling missing data encourages approaches such as multiple imputation or inverse probability weighting, that assume the data are Missing At Random, but also sensitivity analyses that recognise the data may be missing according to the true, unobserved values, that is, Missing Not at Random.   

Future studies should subject the assumptions behind methods for handling non-compliance and missing data to thorough sensitivity analyses. Modern machine learning methods can help reduce reliance on correct model specification. Further research is required to develop flexible methods for handling more complex forms of non-compliance and missing data.  
\end{abstract}


\section{Introduction}

 Health economic studies that evaluate technologies, health policies or public health interventions are required to provide unbiased, efficient estimates of the causal effects of interest. Health economic evaluations are recommended to use data from well-designed randomised controlled trials (RCTs). However, a common problem in RCTs is non-compliance, in that trial participants may depart from their randomised treatment. \citet{Brilleman2015} highlight that when faced with non-compliance, health economic evaluations have resorted to `per protocol' (PP) analyses. A PP analysis excludes those patients who depart from their randomised allocation, and as the decision to switch treatment is likely to be influenced by prognosis, is liable to provide biased estimates of the causal effect of treatment receipt \citep{Ye2014}. 
 
 Another concern is that the required data on outcomes, resource use or covariates may be missing for a substantial proportion of patients. Missing data can arise when patients are lost to follow-up, fail to complete the requisite questionnaires, or when information from routine data sources is incomplete. Most published health economic studies use methods that assume the data are Missing Completely at Random \citep{Noble2012,Leurent2017}. A related concept is `censoring' which refers specifically to when some patients are follow-up for less than the full period of follow-up. For example, in an RCT, those participants who are recruited later to the study, may have their survival data censored at the end of the study period, and it may be reasonable to assume that these data are `censored completely at random' also known as non-informative censoring \citep{Willan2006book}.

More generally in health economic studies, neither non-compliance, nor missing data are likely to be completely at random, and could be associated with the outcome of interest. Unless the underlying mechanisms for the non-compliance or missing data are recognised in the analytical methods, health economic studies may provide biased, inefficient estimates of the causal effects of interest. Appropriate methods for handling non-compliance and missing data have been developed in the wider biostatistics and econometrics literatures (see for example \citet {Angrist1996, Robins1994a, Bang2005, lr02, mk07, Heckman1979}), but reviews report low uptake of these methods in applied health economics studies \citep{Noble2012, Latimer2014, Brilleman2015, Leurent2017}.

Several authors have proposed approaches for handling non-compliance and missing data, suitable for the applied health economics context. In particular, \citet{DiazOrdaz2017}, propose methods for handling non-compliance in cost-effectiveness studies that use RCTs, \citet{Latimer2014} and \citet{White2015} exemplify methods for handling non-compliance with survival (time to event) outcomes, while \citet{Blough2009,Briggs2003,Burton2007,DiazOrdaz2014,DiazOrdaz2016,Faria2014} and \citet{Gomes2013} exemplify methods for handling missing data in health economic evaluations.     

The objective of this chapter is to describe and critique alternative methods for handling non-compliance and missing data in health economics studies. The methods are illustrated with a health economic evaluation that uses data from a single RCT, but we also consider these methods in health economic studies more widely, and identify future research priorities. The chapter proceeds as follows. First, we exemplify the problems of both non-compliance and missing data with the REFLUX case study. Second, we define the necessary assumptions for identifying the parameters of interest, and propose estimation strategies for addressing non-compliance, and then missing data. Third, we contrast some alternative methods in the context of trial-based health economic evaluation. Fourth, we briefly review developments from the wider methodological literature, and identify key research priorities for future health economics studies.

\section{Motivating example: Cost-effectiveness analysis using the REFLUX study}\label{Sec:Reflux}
The REFLUX study was a UK multicentre RCT with a parallel design, in which patients with moderately severe gastro-oesophageal reflux disease (GORD), were randomly assigned to medical management or laparoscopic surgery  \citep{Grant2008, Grant2013}. Resource use and health-related quality of life (QoL), assessed by the EQ-5D (3 levels), were recorded annually for up to five years. Table 1 reports the main characteristics of the study. The original cost-effectiveness analysis (CEA) estimated the linear additive treatment effect on mean costs, $Y_{1i}$  and QALYs, $Y_{2i}$ with a seemingly unrelated regression (SUR) model \citep{Zellner1962, Willan2004}. This model allows for correlation between individual QALYs and costs, and can adjust for different baseline covariates in each of the equations. 

For example, below we have a system of equations that adjust both outcomes for the EQ-5D utility score at baseline, denoted by $\mbox{EQ5D}_{0}$:
\begin{equation}\label{sur}
\begin{array}{c}
Y_{1i} = \beta_{0,1}+ \beta_{1,1} \mbox{treat}_i + \beta_{1,2} \mbox{$\mbox{EQ5D}_{0}$}_{i} + \epsilon_{1i}\\
Y_{2i}= \beta_{0,2}+\beta_{1,2} \mbox{treat}_i + \beta_{2,2} \mbox{$\mbox{EQ5D}_{0}$}_{i} + \epsilon_{2i}
\end{array}
\end{equation}
here $\beta_{1,1}$ and $\beta_{1,2}$ represent the incremental costs and QALYs respectively. The error terms are required to satisfy
$E[\epsilon_{1i}]=E[\epsilon_{2i}]=0, \ 
E[\epsilon_{ \ell i} \epsilon_{\ell'i}] =\sigma_{\ell \ell'},\ 
E[\epsilon_{ \ell i} \epsilon_{\ell'j}]= 0,\ \mbox{\ for\ }\ell,\ \ell'\in\{1,2\}, \mbox{\ and for\ }i\neq j.$ 

 This is an example of a special type of SUR model, in which the  same set of covariates is used in all the equations, the so-called  multivariate regression model. 

Since each equation in the SUR system is a regression, consistent estimation could be achieved by OLS, but would be inefficient. The efficient estimation is generalised least squares (GLS), which requires that the covariance matrix is known \footnote{If we are prepared to assume the errors are bivariate normal, estimation can proceed by maximum likelihood.}, or the feasible version (FGLS), which uses a consistent estimate of the covariance matrix when estimating the coefficients in equation \eqref{sur}  \citep{Zellner1962,Zellner1962a}.
 
There are some situations where GLS (and FGLS) will not be any more efficient than OLS estimation: (1) if  the regressors in a block of equations in the system are a subset of those in another block of the same system, then GLS and  OLS  are identical when estimating the parameters of the smaller set of equations, and (2)
in the special case of multivariate regression, where the SUR  equations have identical explanatory variables, OLS estimation is identical to GLS \citep{Davidson2004}.

The SUR approach can be used to estimate incremental cost and health effects, which in turn can be used to produced incremental cost-effectiveness ratios (ICERs) and incremental net monetary benefits (INBs). Here we focus on the INB, defined as $\mbox{INB}(\lambda) = \lambda \beta_{1,2}-\beta_{1,1}$, where $\lambda$  is the willingness to pay threshold.  The standard error of the INB can be calculated from  the standard errors and correlations of the estimated incremental costs and  QALYs $\hat{\beta}_{1,1}$ and $\hat{\beta}_{1,2}$, following usual rules for estimating the variance of a linear combination of two random variables.

In the REFLUX trial a sizeable minority of the patients crossed over from their randomised treatment assignment (see Table 1), and  the  proportions of patients who switched in the RCT were higher than in routine clinical practice  \citep{Grant2008, Grant2013}. In the RCT, cost and QoL data were missing for a high proportion of the patients randomly assigned to either strategy. \vskip 1ex

\begin{figure}
\caption*{Table 1: The REFLUX study: descriptive statistics and levels of missing data over the five year follow-up period}\centering
\includegraphics[width=1\textwidth]{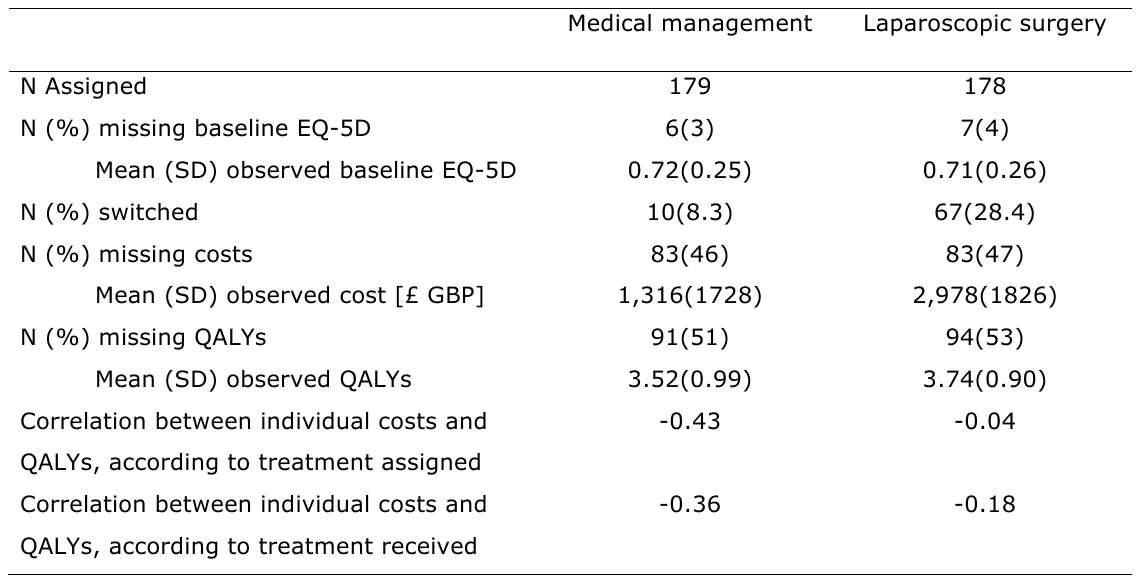}
\end{figure}
The original study reported a PP analysis \citep{Grant2013}, but this is liable to provide a biased estimate of the causal effect of the treatment received. Faced with missing data, the REFLUX study used multiple imputation (MI) and complete case analysis (CCA). In Section  \ref{Sec:ResultsReflux}, we re-analyse the REFLUX study with the methods described in the next sections to report a causal effect of the receipt of surgery versus medical management that uses all the available data, and recognises the joint distribution of costs and QALYs.   

Next, we consider approaches taken to the issue of non-compliance in health economic evaluation more widely, and provide a framework for estimating causal effects in CEA.

\section{Non-compliance in health economic evaluation}
The intention-to-treat  (ITT) estimand can provide an unbiased estimate of the intention to receive a particular treatment, but not of the causal effect of the treatment actually received. Instrumental variable (IV) methods can identify the complier average causal effect (CACE), also known as the local average treatment effect (LATE). In RCTs with non-compliance, random assignment can act as the instrument for treatment receipt, provided it meets the IV criteria for identification \citep{Angrist1996}. An established approach to IV estimation is two-stage least squares (2sls), which provides consistent estimates of the CACE when the outcome model is linear, and non-compliance is binary \citep{Baiocchi2014}.

In CEA that use RCT data, there is interest in estimands, such as the relative cost-effectiveness for compliers. An estimate of the causal effect of the treatment received can help with the interpretation of results from RCTs with different levels of non-compliance to those in the target population. The causal effect of treatment receipt is also of interest in RCTs with encouragement designs, for example of behavioural interventions to encourage uptake of new vaccines \citep{Duflo2008}, and for trial designs, common in oncology, which allow treatment switching according to a variable time period, such as after disease progression \citep{Latimer2014} (see Section \ref{subsec:survival}).  

In CEA,  methods that report the causal effect of the treatment have received little attention. \citet{Brilleman2015} found that most studies that acknowledged the problem of non-compliance reported PP analyses, while \citet{Hughes2016} suggested that further methodological development was required. The context of trial-based CEA raises important complexities for estimating CACEs that arise more generally in studies with multivariate outcomes. Here to provide accurate measures of the uncertainty surrounding a composite measure of interest, for example the INB, it is necessary to recognise the correlation between the endpoints, in this case, cost and health outcomes \citep{Willan2003, Willan2006a}.  

We now provide a framework for identifying and estimating the causal effects of the treatment received. First, we present the three stage least squares (3sls) method \citep{Zellner1963}, which allows the estimation of a system of simultaneous equations with endogenous regressors. Next, we consider a bivariate Bayesian approach, whereby the outcome variables and the treatment received are jointly modelled as dependent on random assignment. This extends IV unrestricted reduced form  \citep{Kleibergen2003}, to the setting with multivariate outcomes.

\section{Complier Average Causal effects with bivariate outcomes}\label{Sec:CACE}

We begin by defining more formally our estimand and assumptions. Let $\Yc$ and $\Ye$ be the continuous bivariate outcomes, and $Z_i$ and $D_i$ the binary random treatment allocation and treatment received respectively, corresponding to the $i$-th individual. The bivariate endpoints $Y_{1i}$ and $Y_{2i}$ belong to the same individual $i$, and thus are correlated. We assume that there is an unobserved confounder $U$, which is associated with the treatment received and either or both of the outcomes.
We assume that the \textbf{(i) Stable Unit Treatment Value Assumption (SUTVA)} holds:
the potential outcomes of the $i$-th individual are unrelated to the treatment status of all other individuals (known as \textit{no interference}), and that for those who actually received treatment level $z$, their observed outcome is the potential outcome corresponding to that level of treatment.

Under SUTVA, we can write the potential  treatment received by the $i$-th subject  under the  random assignment at level $z_i \in \{0, 1\}$ as  $D_i\left(z_i\right)$. Similarly, $Y_{\ell i}\left(z_i,d_i\right)$ with $\ell\in\{1,2\}$ denotes the corresponding potential outcome for endpoint $\ell$, if the $i$-th subject were allocated to level $z_i$ of the treatment and received level $d_i$. There are four potential outcomes. Since each subject is randomised to one level of treatment, only one of the potential outcomes per endpoint $\ell$, is observed, \ie $Y_{\ell i}=Y_{\ell i}(z_i,D_i(z_i))=Y_i(z_i)$.

The CACE for outcome $\ell$ can now be defined as
\begin{equation}\label{cace}
\theta_{\ell}=E\left[\{Y_{\ell i}(1)-Y_{\ell i}(0)\}\big|\{D_i(1)-D_i(0)=1\}\right].
\end{equation}

In addition to SUTVA,  the following  assumptions are sufficient for identification of the CACE, \citep{Angrist1996}: \vskip 1ex
\noindent\textbf{(ii) Ignorability of the treatment assignment}: $Z_i$ is independent of unmeasured confounders (conditional on measured covariates)  and the potential outcomes $Z_{i} \ci U_i,D_{i}(0),D_{i}(1),Y_{i}(0),Y_{i}(1).$
\\ \textbf{(iii) The random assignment predicts treatment received}: $Pr\{D_i(1)=1\}\neq Pr\{D_i(0)=1\}$.
\\  \textbf{(iv) Exclusion restriction}: The effect of $Z$ on $Y_\ell$ must be via an effect of $Z$ on $D$; $Z$ cannot affect $Y_\ell$ directly. \\ \textbf{(v) Monotonicity:} $D_i(1)\geq D_i(0)$.

 The CACE can now be identified from equation \eqref{cace} without any further assumptions about the unobserved confounder; in fact, $U$ can be an effect modifier of the relationship between $D$ and $Y$ \citep{Didelez2010}.

In the REFLUX study, the assumptions concerning the random assignment, (ii) and (iii), are justified by design. The exclusion restriction assumption is plausible for the cost endpoint, as the costs of surgery are only incurred if the patient actually has the procedure, and it seems unlikely that assignment rather than receipt of surgery would have a direct effect on QALYs. The monotonicity assumption rules out the presence of defiers, and it seems reasonable to assume that there are no trial participants who would only receive surgery if they were randomised to medical management, and vice versa.   Equation \eqref{cace} implicitly assumes that receiving the intervention has the same average effect on the linear scale, regardless of the level of $Z$ and $U$.  

Since random allocation, $Z$, satisfies assumptions (ii)-(iv), we say it is an instrument for $D$. For a binary instrument, the simplest approach to  estimate equation \eqref{cace} within the IV framework is the Wald estimand \citep{Angrist1996}:
\[
{\beta}_{\ell} = \frac{E(Y|Z=1) - E(Y|Z=0)}{E(D|Z=1) - E(D|Z=0)}
\]
Typically, estimation of these conditional expectations proceeds via the so-called two-stage least squares (2sls). The first stage is a linear regression model that estimates the effect of the instrument on the exposure of interest, here treatment received on treatment assigned. The second stage is the outcome regression model, but fitted on the predicted treatment received from the stage one regression:
\begin{eqnarray}\label{2sls}
\nonumber D_i &= &\alpha_0 + \alpha_1Z_i + \epsilon_{0,i}\\
Y_{\ell i} &=& \beta_0 + \beta_{1, \ell} \widehat{D_i} + \epsilon_{\ell,i}
\end{eqnarray}
\noindent where $\widehat{\beta}_{1, \ell} $ is an estimator for $\beta_{\ell}$. Covariates can be included in both model stages. 

For 2sls to be consistent, the first stage model  must  be the parametric linear regression implied by the second stage, that is, it must include all  the covariates and interactions that appear in the second stage model \citep{Wooldridge2010}. 

The asymptotic standard error for the 2sls estimate of the CACE is given in \citet{Imbens1994}, and is available in commonly used software packages. However, 2sls can only be  readily applied to univariate outcomes, which raises an issue for CEA as ignoring the correlation between the two endpoints would provide inaccurate measures of uncertainty. A simple way to address this problem would be to apply 2sls directly within a net benefit regression approach \citep{Hoch2006}. However, it is known that net benefit regression is very sensitive to outliers, and to distributional assumptions \citep{Willan2004}, \citep{Mantopoulos2016}. Instead, we focus on strategies for jointly estimating the CACE on QALYs and costs. The first approach combines SURs (equation \ref{sur}) with 2sls (equation \ref{2sls}) to obtain CACEs for both outcomes accounting for their correlation, and is known as three-stage least squares (3sls). The second is a Bayesian estimation method for the system of simultaneous equations.

\subsection{Three-stage least squares}
Three-stage least squares (3sls) was developed as a generalisation of 2sls for systems of equations  with endogenous regressors,  \ie any explanatory variables which are correlated with the error term in its corresponding equation  \citep{Zellner1963}. 
\begin{eqnarray}\label{3sls}
\nonumber D_i &=& \alpha_0 + \alpha_1Z_i + e_{0i}\\
 Y_{1i} &=& \beta_{01} + \beta_{IV,1} {D_i} + e_{1i} \\
\label{3sls2}Y_{2i} &=& \beta_{02} + \beta_{IV,2} {D_i} + e_{2i}
\end{eqnarray}
As with 2sls, the models can be extended to include baseline covariates. 

All the parameters appearing in the system of equations \eqref{3sls} and \eqref{3sls2} are estimated jointly. Firstly, the IV model is estimated ``outcome by outcome'', for example by applying 2sls. This will be consistent but inefficient. The residuals from this 2sls models, that is $e_{1i}$ and $e_{2i}$, can be now used to estimate the covariance matrix that relates the outcome models. This is similar to the step used on a SUR with exogenous regressors (equation \eqref{sur}) for estimating the covariance matrix of the error terms from the two equations \eqref{3sls} and \eqref{3sls2}. 
This estimated covariance matrix is used when solving the estimating equations formed by stacking the equations vertically \citep{Davidson2004}. 

Provided that the identification assumptions (i)-(v) are satisfied,
Z is independent of the residuals at each stage, i.e.  $Z \ci e_{0i}$, $Z \ci e_{1i}$, and $Z \ci e_{2i}$, the estimating equations can be solved by FGLS which avoids distributional assumptions, and is also robust to heteroscedasticity of the errors across the different linear models for the outcomes \citep{Greene2002}.  As the 3sls method uses an estimated covariance matrix, it is only asymptotically efficient \citep{Greene2002}.
  If the error terms in each equation of the system and the instrument are not independent, the 3sls estimator based on FGLS is not  consistent, and other estimation approaches, such as generalised methods of moments (GMM) warrant consideration   \citep{Schmidt1990}. 

In the just-identified case, that is when there are as many endogenous regressors as there are instruments,  classical theory shows that the GMM and the FGLS estimators coincide \citep{Greene2002}. 

\subsection{Bayesian estimators}

  \citet{Nixon2005} propose Bayesian bivariate models for the expectations of the two outcomes (e.g. costs and QALYs) in CEA, which have a natural appeal for this context as each endpoint can be specified as having a different distribution. The parameters in the models are simultaneously estimated, allowing for proper Bayesian feedback and propagation of uncertainty. On the other hand, univariate instrumental variables models within the Bayesian framework have been previously developed \citep{Burgess2012,Kleibergen2003, Lancaster2004}. 

This method recasts  the first and second stage equations familiar from 2sls, eqs. \eqref{2sls}, in terms of a recursive equation model,  which can be ``solved'' by substituting the parameters of the first into the second. Such solved  system of equations is called the \textit{reduced form}, and it expresses explicitly  how the endogenous variable $D$ and the outcome $Y_{\ell }$ jointly  depend on the instrument.
\begin{eqnarray}\label{reduced}
\nonumber D &= &\alpha_0 + \alpha_1Z+ \nu_{0}\\
Y_{\ell } &=& \beta_0^* + \beta_{1, \ell} \alpha_1Z + \nu_{\ell}
\end{eqnarray}
\noindent where $\beta_0^*=\beta_0+  \beta_{1, \ell}\alpha_0$,  $\nu_{0}=\epsilon_0$ and $\nu_{\ell}= \epsilon_\ell + \beta\epsilon_0$.

The parameter of interest $\beta_{1, \ell}$ is identified, since by the IV assumptions, $\alpha_1\neq 0$.

The extension  of this reduced form to multivariate outcomes proceeds as follows.  Let $(D_{i}, Y_{1i}, Y_{2i})^\top$ be the transpose of the vector of outcomes, which now includes the endogeneous variable $D$ as well as the bivariate endpoints of interest.

The reduced form can now be written in terms of the linear predictors of $D_{i}, Y_{1i}, Y_{2i}$ as :
\begin{equation}
\begin{array}{l}
\mu_{0i} =\beta_{0,0} + \beta_{1,0}Z_i \\
\mu_{1i} = \beta_{0,1}+ \beta_{1,1} \beta_{1,0}Z_i  \\
\mu_{2i}=  \beta_{0,2}+\beta_{1,2}\beta_{1,0}Z_i
\end{array}
 \end{equation}
with $\beta_{0,0}=\alpha_0$,  $\beta_{1,0}=\alpha_1$.

We treat $D_{i}, Y_{1i}, Y_{2i}$ as multivariate normally distributed, so that:
\begin{equation}
\begin{array}{l}
\left(\!\!\begin{array}{c}
D_{i}\\ Y_{1i} \\ Y_{2i}\end{array}\!\!\right)
\sim N\left\{\!\!\left(\begin{array}{c}
\mu_{0i}\\
\mu_{1i}\\
\mu_{2i}\end{array}\right),  \Sigma= \left(\!\!\!\begin{array}{ccc} \s_{0} & s_{01} & s_{02} \\  s_{01}& \s_{1} & s_{12} \\
 s_{02}& s_{12} & \s_{2}
 \end{array} \!\!\!\right)
\!\!\right\};
\end{array}
\end{equation}
where $s_{ij}=\mbox{cov}(Y_i,Y_j)$, and the causal treatment effect estimates are $\beta_{1,1}$ and $\beta_{1,2}$ respectively.  For the implementation, we use vague normal priors for the regression coefficients, \ie $\beta_{m,j}\sim N(0, 10^2)$, for $j \in\{0,1,2\}, m\in\{0,1\}$, and a Wishart prior for the inverse of $\Sigma$ \citep{GelmanHill}.

\subsection{Comparison of  Bayesian versus 3sls estimators for CEA with non-compliance}

The performance of 3sls and Bayesian methods for obtaining compliance-adjusted estimates was found to be similar in a simulation study \citep{DiazOrdaz2017}. Under the IV and monotonicity assumptions, both approaches performed well in terms of bias and confidence interval coverage, though the Bayesian estimator reported wide CIs around the estimated INB in small sample size settings. The 3sls estimator reported low levels of bias and good CI coverage throughout.

These estimators rely on a valid IV and for observational data settings, the assumptions required for identification warrant particular scrutiny. While the use of IV estimators in health economics studies has not been extensive, the development of new large, linked observational datasets offer new opportunities for harnessing IVs to estimate the causal effects of treatments received. In particular, areas such as Mendelian randomisation offer the possibility of providing valid instruments and have previously been used with the Bayesian estimators described above \citep{Burgess2012}.

Methods are also available which do not rely on an IV, but still attempt to estimate a causal effect of treatment receipt. We briefly review one of these methods below.

\subsection{Inverse probability weighting for non-compliance}
Inverse probability weighting  (IPW) can also be used to obtained compliance-adjusted  estimates. Under IPW, observations that deviate from protocol are censored, similar to a PP analysis. To avoid selection bias, the data from those participants that continue with the treatment protocol are weighted, to represent the complete (uncensored) sample according to observed characteristics. 
The weights are given by the inverse of the probability of complying, conditional on the covariates included in the \textit{non-compliance model}.  

The target estimand is the causal average treatment effect (ATE). For IPW to provide unbiased estimates of the ATE, requires that the model includes all baseline and time-dependent variables that predict both treatment non-adherence and outcomes. This is often referred to as the ``no unobserved confounder'' assumption \citep{Robins2000}.  \citet{Latimer2014} illustrate IPW for health technology assessment. 

The IPW method cannot be used when there are covariates values that perfectly predict treatment non-adherence, that is when there are covariate levels where the probability of non-adherence is equal to one \citep{Robins2000a, Hernan2001, Yamaguchi2004}. We consider further methods for handling non-compliance in health economic studies in Section \ref{Sec:FurtherTopics}. We turn now to the problem of missing data.      

\section{Missing data}\label{Sec:missdata}

The approach to handling missing data should be in keeping with the general aim of providing consistent estimates of the causal effect of the interventions of interest. However, most published health economic evaluations simply discard the observations with missing data, and undertake complete case analyses (CCA) \citep{Noble2012, Leurent2017}. This approach is valid, even if inefficient, where the probability of missingness is independent of the outcome of interest given the covariates in the analysis model, that is there is covariate-dependent missingness (CDM)  \citep{White2010}. A CCA is also valid when the data are missing completely at random (MCAR) \citep{lr02}, that is the missing data do not depend on any observed, or unobserved value. Similarly, when data are censored for administrative reasons unrelated to the outcome, then they may be assumed to be censored completely at random \citep{Willan2006book}.   

In many health economic evaluations it is more realistic to assume that the data are missing (or censored) at random (MAR), that is, that the probability that the data are observed, only depends on observed data \citep{lr02}.  For example, a likelihood-based approach that only uses the observed data, can still provide valid inferences under the MAR assumption, if the analysis adjusts for all those variables associated with the probability of missing data \citep{mk07} 

For SUR systems, it can often be the case that each equation has different number of observations, either because of differential missingness in the outcomes or because each equation included different regressors, which in turn have alternative missing data mechanisms. 
Estimation methods for  SUR models with unequal numbers of observations  have been considered by \citet{ Schmidt1977}.  Such estimates would then be valid under the asumptions of CDM or MCAR. In addition, if the estimation method was likelihood-based (MLE, Expectation-Maximisation and Bayesian), the estimates are also  valid under MAR, again provided the models adjust for all the variables driving the missingness. Estimation of SUR systems with unequal number of observations based on Maximum-likelihood was presented  by \citet{Foschi2002} within the frequentist literature. Bayesian likelihood estimation of SURs with unequal number of observations was developed by \citet{Swamy1975}.

In general, MI, IPW and full Bayesian methods are the recommended approaches if the data can be assumed to be MAR \citep{Rubin1987}. Methods that assume data are MAR have been proposed in health economics, for example to handle missing resource use and outcomes \citep{Briggs2003}  or unit costs \citep{Grieve2010}. In the specific context of censored data, time to event parametric models have been adapted to the health economics context, and assume that the censoring is un-informative, that is the probability that the data are censored only depends on the observed data, see for example \citet{Lin1997, Carides2000, Raikou2004, Raikou2006}, and also section 7.3.3.     

An alternative assumption is that the missing values are associated with data that are not observed, that is the data are missing not at random (MNAR). Methods guidance for the analysis of missing data recommends that, while the primary or base case analysis should present results assuming MAR, sensitivity analyses should be undertaken that allow for the data to be MNAR \citep{Sterne2009}. The review by \citet{Leurent2017} highlights that there are very few examples in health economics studies that consider MNAR mechanisms (see section \ref{Sec:sensMAR}).

\subsection{Multiple Imputation (MI)}\label{sec:missing}
MI is a  principled tool for handling missing data \citep{Rubin1987}. 
MI requires the analyst to distinguish between two statistical models. The first
model called the \textit{substantive model}, \textit{model of interest} or \textit{analysis model} is the one that would have been used had the data been complete. The second model, called the
{\it imputation model},  is used to describe the conditional distribution of the missing data, given the observed, and must include the outcome. 
Missing data are imputed with the imputation model, to produce several completed data sets. Each set is then analysed separately using the original analysis model, with the resultant parameter estimates and associated measures of precision combined by Rubin's formulae \citep{Rubin1987}, to produce the MI estimators and their variances. Under the MAR assumption, this will produce consistent estimators \citep{lr02,schafer97}.
 MI can increase precision, by incorporating partly observed data, and remove potential bias from undertaking a CCA when data are MAR. If only outcomes are missing, MI may not improve upon a conventional likelihood analysis  \citep{White2010}. Nevertheless, MI can offer improvements if the imputation model includes variables that predict missingness and outcome, additional to those in the analysis model. The inclusion of these so-called auxiliary variables can make the assumption that the data are MAR more plausible.

A popular approach to MI is full-conditional specification (FCS) or {\em chained equations} MI, where draws from the joint distribution are approximated using a sampler consisting of a set of univariate models for each incomplete variable, conditional on all the other variables \citep{vb12}.
FCS has been adapted to include interactions  and other non-linear terms in the imputation model, so that the imputation model contains the analysis model. This approach is known as substantive model compatible FCS (SMCFCS) \citep{Bartlett2014}. 

 \citet{Faria2014} present a guide to handling missing data in CEA conducted within trials, and show how MI can be implemented in this context.

\subsection{IPW for valid inferences with  missing data}
IPW can address missing data,  by using  weights  to re-balance complete cases so that they represent the original sample. The use of IPW for missingness is similar to re-weighting different sampling fractions within a survey.
The units with a low probability of being fully observed, are given a relatively high weight so that they represent units with similar (baseline) characteristics who were not fully observed, and would be excluded from a CCA.

The weighting model $p(R_i=1\vert X_i, Y_i) $ is called the \textit{probability of missingness model}  (POM). From this model, we can estimate the probability of being observed, by for example, fitting a logistic model  and obtaining fitted probabilities $\widehat{\pi_i}$. The weights $w_i$, are then the inverse of these estimated probabilities, i.e. $w_i=\frac{1}{\widehat{\pi_i}}$.

The IPW approach incorporates this re-weighting in applying the substantive model to the complete cases. IPW provides consistent estimators when the data are MAR and the POM models are correctly specified. The variance of the IPW estimator is consistently estimated provided the weighting is taken into account, by for example using a sandwich estimator \citep{Robins1994}. IPW is simple to implement when there is only one variable with missing data, and the POM model only includes predictors that are fully-observed. It is still fairly straight-forward, when there are missing data for more variables but the missing data pattern or missingness is \textit{monotone}, which is the case when patients drop out of the study follow-up after a particular timepoint. The tutorial by \cite{Seaman2011} provides further detail on using IPW for handling missing data. 

IPW has rarely been used to address missing data in health economic evaluation  \citep{Leurent2017, Noble2012}. Here a particular concern is that there may be poor overlap between those observations with fully- versus partially- observed data according to the variables that are in the POM model, leading to unstable weights.  There are precedents for using IPW in health economics settings with larger datasets, and where lack of overlap is less of a problem (see \citet{Jones2006}). IPW and MI both provide consistent estimates under MAR, if either the POM or imputation models are correctly specified. However, MI is often preferred to IPW, as it is usually more efficient and more flexible approach for handling non-monotone patterns of missing data, for example when data are missing at intermittent time-points during the study follow-up.

\subsection{Bayesian analyses}

Bayesian analyses naturally distinguish between  observed data and  unobserved quantities. All unobserved quantities are viewed as unknown ``parameters'' with an associated probability distribution. From  this perspective, missing values simply become extra parameters to model and obtain posterior distributions for. Let $\mathbf{\theta}$ denote the parameters of the full data  model, $p(y,r\vert x; \mathbf{\theta})$. This model can be factorised as the substantive model times the POM, i.e. the model for the missingness mechanism 
$p(y\vert x; \gamma)p(r\vert y, x; \psi)$.

For Bayesian ignorability to hold, we need to assume MAR, and in addition that $\gamma$ and $\psi$ are a distinct set of parameters, i.e. variation independent, with the prior distribution for $\mathbf{\theta}$ factorising into 
$p(\mathbf{\theta})= p(\gamma)p(\psi)$.

This means that under Bayesian ignorability, if there are  covariates in the analysis model which have missing data,  additional models for the distribution of these covariates are required. These models often involved  parametric assumptions about the full data. 
The Bayesian approach also uses priors to account for uncertainty about the POM. This is in contrast to IPW, for example, which ignores the uncertainty in the parametric POM, and relies on this model being correctly specified.  It is possible to use auxiliary variables under Bayesian ignorability. Full details are provided by \cite{Daniels2008}.

\section{Application to estimating the CACE with non-response in the REFLUX study}\label{Sec:ResultsReflux}

We now apply 3sls and Bayesian IV models to obtain CACE of the surgery vs medical management treatment, using  the REFLUX example.  We compare CCA with MI, IPW and full Bayesian approaches assuming the missing data under MAR. Only  48\% of trial participants had completely observed costs and QoL data, with a further 13  missing baseline $\mbox{EQ5D}_{0}$. 

 For the CCA, SUR is applied to the cases with full information to report the INB according to the ITT and PP estimands. As Table 2 shows, the PP estimate is somewhat lower than the ITT and CACE estimates, reflecting that those patients who switch from surgery to medical management and are excluded from this analysis, have a somewhat better prognosis than those who follow their assigned treatment. Assuming that randomisation is a valid IV and that monotonicity holds, either 2sls or Bayesian methods provide  CACE estimates  of the INB for compliers. However, these CCA assume that the missingness is independent of the outcomes, given the treatment received and the baseline EQ5D, that is there is covariate-dependent missingness (CDM).  This assumption may be implausible. We now consider strategies valid under a MAR assumption, specifically, we use MI and IPW coupled with 3sls, and a full Bayesian analysis, to obtain valid inferences for the CACE, that use all the available data.

For the MI,  we  considered including auxiliary variables in the imputation model, but none of the additional covariates  were associated with the missingness and the value of the variables to be imputed. We imputed  total cost, QALYs and baseline $\mbox{EQ5D}_{0}$, 50 times by FCS, with predictive mean matching  (PMM), taking the five nearest neighbours as donors  \citep{White2011}. The imputation models must contain, all the variables in the analyses models, and so we included  treatment receipt in the imputation model, and stratified by randomised arm.  We then applied 3sls to each of the 50 MI sets, and combined the results with Rubin's formulae \citep{Rubin1987}.

The IPW required POM models for the baseline EQ5D, cost, and QALY  respectively. Let $R_{0i}$, $R_{1i}$  and $R_{2i}$ be the respective missingness indicators. The missingness pattern is almost  monotone, with 156 individuals with observed  $\mbox{EQ5D}_{0}$, i.e.  $R_{0i}=1$, then a further   16 with $R_{0i}=R_{1i}=1$, and  10 with all $R_{ji}=0$.\footnote{For simplicity we enforced a monotone pattern of missingness by excluding the three individuals with missing $\mbox{EQ5D}_{0}$ but observed costs, i.e.  $R_{0i}=0$ but $R_{1i}=1$.} 

With this monotone missing pattern we have POMs for: $R_{0i}$ on all other fully observed baseline covariates; $R_{1i}$ on fully observed baseline covariates, randomised arm treatment receipt, $R_0$ and  $\mbox{EQ5D}_{0}$; and $R_{2i}$
 on fully observed baseline covariates, randomised arm, treatment receipt, $R_0$, $\mbox{EQ5D}_{0}$, $R_1$, and cost.  We fitted these models with logistic regression, used backward stepwise selection, and only kept those regressors with p-values less than 0.1.
This resulted in POMs as follows: an empty model for  $R_{0i}$; age,  randomised group, treatment received, and $\mbox{EQ5D}_{0}$, for the $R_{1i}$, and only $\mbox{EQ5D}_{0}$ for $R_{2i}$.
We obtained the predicted probabilities of being observed, and weighted the complete cases in the 3sls analysis by the inverse of the product of these three probabilities. 

In the Bayesian analyses, to provide a posterior distribution for the missing values, a model for the distribution  of baseline $\mbox{EQ5D}_{0}$ was added  to the treatment received and outcome models. Bayesian posterior distributions of the mean costs, QALYs and INB  were then summarised by their median value and 95\% credible intervals.  
Table 2 shows that the CACE estimates under MAR are broadly similar to those from the CCA, but in this example the MI and Bayesian approaches provided estimates with wider CI. In general, MI can result in more precise estimates than CCA, when the imputation model does include auxiliary variables.  The general conclusion of the study is that  surgery is relatively cost-effective in those patients with GORD who are compliers. The results are robust to a range of approaches for handling the non-compliance and missing data. 
 
\begin{figure}
\caption*{Table 2: The REFLUX study: cost-effectiveness results according to estimand, estimator and approach taken to missing data} \label{Table1} \centering
\includegraphics[width=1\textwidth]{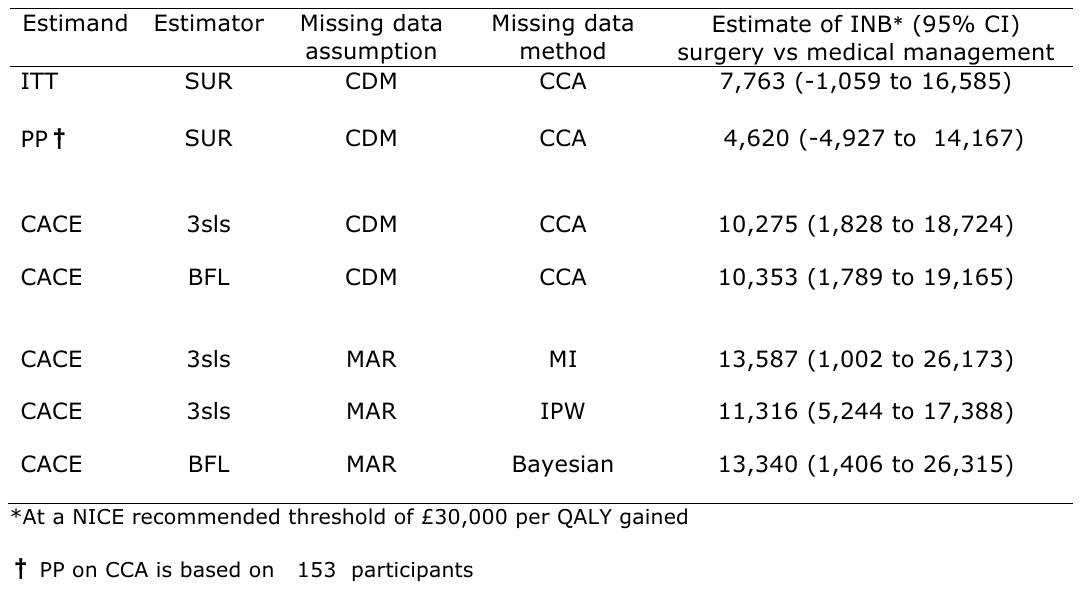}
\end{figure}

\section{Further topics}\label{Sec:FurtherTopics}

We now consider further methodological topics relevant to health economic studies faced with non-compliance and missing data.  First, we discuss sensitivity analyses for departures from the IV assumptions  (section \ref{Sec:sensIV}), and  the MAR assumption  (section \ref{Sec:sensMAR}). We also consider approaches for addressing non-compliance and missing data for non-continuous outcomes (section \ref{Sec:nonnormal}), and clustered data (section \ref{Sec:clus}) before offering conclusions and key topics for further research (section \ref{Sec:discussion}).   

\subsection{Sensitivity analyses to departures from the IV assumptions to adjust for non-compliance}\label{Sec:sensIV}
The preceding sections detailed how an instrumental variable for the endogeneous exposure of interest can be used to provide consistent estimates of the CACE as long as particular assumptions are met. However, it is helpful to consider whether conclusions are robust to departures from these assumptions, in particular in those settings where these underlying assumptions might be less plausible. For example, sensitivity to the exclusion restriction assumption, can be explored by extending the Bayesian model described. The analyst can specify priors on the non-zero direct effect of the IV on the outcome \citep{Conley2012, Hirano2000}. Here it must be recognised that as the models are only weakly identified,  the results may be highly sensitive to the parametric choices for the  likelihood and the  prior distributions.  Within the frequentist IV framework, sensitivity analyses can build on 3sls to consider potential violations of the exclusion restriction and monotonicity assumptions. See \citet{Baiocchi2014} for a tutorial. 
Another important assumption is that the IV strongly predicts treatment receipt, and this might well be satisfied in most clinical trials \citep{Zhang2014}. In health economics studies that use RCTs with encouragement designs, or Mendelian randomisation observational studies where the IV is a gene, the IV may well be weak, and this can lead to biased estimates. Here, Bayesian IV methods have been shown to perform better than 2sls methods  \citep{Burgess2012}. 

 \subsection{Sensitivity analyses to departures from MAR assumptions}\label{Sec:sensMAR}
There has been much progress in the general econometrics and biostatistics literature on developing methods for handling data that are assumed MNAR (see for example \citet{Heckman1976,  Rubin1987, ck13}), but there are few examples in the applied health economics literature \citep{Noble2012, Faria2014, Leurent2017, Jones2006} 
For settings when data are assumed to be MNAR, the two main groups of methods proposed are: selection models and pattern mixture models. Selection models postulate a mechanism by which the data are observed or  `selected', according to the true, underlying values of the outcome variable \citep{Heckman1979}. By contrast pattern-mixture models specify alternative distributions for the data according to whether or not they are observed \citep{lr02,ck13}.

Heckman selection models have been used in settings with missing survey responses particularly in HIV studies (see for example \citet{Clark2014}).
Heckman's original proposal was a two-step estimator, with a binary choice model to estimate the probability of observing the outcome in the first stage, with the second stage then using the estimates in a linear regression to model the outcomes \citep{Heckman1976}.  Recent extensions allow for other forms of outcomes, such as binary measures, by using GLMs for the outcome equation \citep{Marra2013}, and recognise the importance of appropriate model specification and distributional assumptions \citep{Vella1998, Das2003, Marchenko2012, McGovern2015}. The Heckman selection model requires the selection model to include covariates not in the outcome model, to avoid collinearity issues. This relies on an important untestable assumption which tends to remain hidden: these variables must meet the criteria for the exclusion restriction, that is, they must predict missingness, and also be conditionally independent of the outcome \citep{Puhani2000}. 

Pattern mixture models (PMM) have been advocated under MNAR as they make more accessible, transparent assumptions about the missing data mechanism \citep{Molenberghs2014, ck13}. The essence of the pattern-mixture model approach under MNAR is that it recognises that the conditional distribution of partially observed variables, given the fully observed variables, may differ between units that do, and do not, have fully observed data \citep{ck13}. Hence, a PMM approach allows the statistics of interest to be calculated across the units with observed data (pattern 1), and then for those with missing data (pattern 2). For the units with missing data, offsets terms are added to the statistics of interest calculated from the observed data (pattern 2). The offset terms, also known as sensitivity parameters, can differ according to the treatment groups defined according to treatment assigned, or if there is non compliance according to the treatment received.  
MI is a convenient way to perform PMM sensitivity analyses. The offset method can be easily implemented for variables which have been imputed under linear regression models. The software package SAS has implemented PMM for other models within the MI procedure \texttt{MI}.

As \citet{Leurent2017} highlight in the context of health technology assessment, few applied studies have used PMM. One potential difficulty is that studies may have little rationale for choosing values for the sensitivity parameters \citep{Leurent2018}. The general methodological literature has advocated expert opinion to estimate the sensitivity parameters, which can then be used as informative priors within a fully Bayesian approach \citep{White2007}. Progress has been made by \citet{Mason2017} who developed a tool for eliciting expert opinion about missing outcomes for patients with missing versus observed data, and then used these values in a fully Bayesian analysis. In a subsequent paper \citet{Mason2018} extend this approach to propose a framework for sensitivity analyses in CEA for when data are liable to be MNAR.

\subsection{Other type of outcomes}\label{Sec:nonnormal} 
 The methods described have direct application to health economics studies with continuous endpoints such as costs and health outcomes such as QALYs. We now briefly describe extensions to settings with other forms of outcomes, namely binary (e.g. admission to hospital) or time to event (duration or survival) endpoints.

\subsubsection{Binary outcomes}

 Often, researchers are interested in estimating the causal treatment effect on a binary outcome. The  standard 2sls estimator requires that both  stages to be  linear. This is often a good estimator for the risk difference,  but may result in estimates that do not respect the bounds of the probabilities, which must lie between 0 and 1.  Several alternative two-stage  estimators have been proposed. 
When odd ratios are of interest,  two IV methods  based on plug-in estimators have been proposed \citep{Terza2008}. The first-stage of these approaches is the same, a linear model of the endogenous regressor $D$ on the instrument $Z$ (and the covariates $X$ if using any in the logistic model for the outcome). 
Where they differ is in the outcome model, the second stage.  
The first strategy, so-called  `standard' IV estimator or two-stage predictor substitution (2SPS)  regression estimates the causal log odds ratio with the coefficient for the fitted $\hat{D}$. However, this will be biased for the conditional odds ratio of interest, with the bias increasing  with the strength of the association between $D$ and $Y$ given $Z$. \citep{Vansteelandt2011}.

The second strategy, has been called  `adjusted two-stage estimate' or two-stage residual inclusion (2SRI). The second-stage equation, i.e. the outcome model,  fits a logistic regression of $Y$ on $D$ and the residual from the first-stage regression (as well as other exogeneous baseline covariates $X$, if using any).  However, the non-collapsibility of logistic regression means that 2SRI only provides asymptotically unbiased estimates of the CACE if there is no unmeasured confounding, i.e. when the model includes all the covariates that predict non-compliance and the outcomes \citep{Cai2011}. This is because when there is unobserved confounding, the estimate obtained by 2SRI is conditional on this unobserved confounding, and since it is unobserved, cannot be compared in any useful way with the population odds ratio of interest. If we had measured those confounder, we could marginalise over their distribution to obtain the population (marginal) odds ratio \citep{Burgess2013}.  See  \citet{Clarke2012} for a comprehensive review of IV methods for binary outcomes. 

  \citet{Chib2002} developed simultaneous probit models, involving latent normal variables for both the endogenous discrete regressor and the discrete dependent variable from the Bayesian perspective, while \citet{Stratmann1992} developed the full maximum likelihood version. 

The MI approaches previously described for handling missing data under MAR and MNAR can all be applied to binary outcomes, for example by using a logistic imputation model within FCS MI  \citep{lr02,ck13}. The IPW strategies proceed for binary outcomes in the same way as for continuous outcomes.
 
\vskip 2ex
\subsubsection{Time to event data} \label{subsec:survival}
The effect of treatment on time to event outcomes, is often of interest for example in oncology trials, or in evaluations of behavioural interventions, where time to stop particular behaviour (e.g. time to smoking cessation) are often reported. 

A common IV method for estimating the causal effect of treatment receipt on survival outcomes is the so-called rank preserving structural  failure time models (RPSFT) \citep{Robins1991}, later extended to account for censoring by \citet{White1999} and applied to health technology assessment by \citet{Latimer2014a, Latimer2014}.
These models are called rank preserving, because it is assumed that if subject $i$ has the  event before subject $j$ when both received the treatment, then subject $i$ would also have a shorter failure time if neither received the treatment i.e. $T_i(a)<T_j(a)$,  for $a\in\{0,1\}$. This is, randomisation is assumed to be an IV, and therefore to meet the exclusion restriction such that the
counterfactual survival times are assumed independent of the randomised arm.

The target estimand is the average treatment effect amongst those who follow the same treatment regime, i.e the average treatment effect on the treated (ATT), as opposed to a CACE.  
In the simplest case, where treatment is a one-off, all-or-nothing (e.g. surgery), let $\lambda_{T(1)}(t)$ be the hazard of the subjects who received the treatment. 

We can use the g-estimation \citep{Robins2000b} procedure to obtain the treatment-free hazard function $\lambda_{T(0)}(t)$ to obtain 
 $\lambda_{T(1)}(t)= \lambda_{T(0)}(t)e^{-\beta}$. We refer to $\beta$ as the causal rate ratio, as treatment multiplies the treatment-free hazard by a $e^{-\beta}$. 
 As this is an IV method, it assumes that $Z$ is a valid instrument, however, instead of monotonicity, it requires that treatment effect, expressed as $\beta$, is the same for both randomised arms, often referred to as \textit{no effect modification by $Z$}, which may be plausible in RCTs that are double-blinded. This is a very strong assumption if, for example, the group randomised to the control regimen, receive the treatment later in the disease pathway (e.g. post progression), compared to the group randomised to treatment.  
This method has also been adapted for time-updated treatment exposures.
\subsubsection{Missing data in time to event analyses}

 For survival analysis, the outcome consists of a variable $T$ representing time to the event of interest and an event indicator $Y$. If $Y_i=1$, then that individual had the event of interest at time $T_i$, but if $Y_i=0$, then $T_i$ records the censoring time,  the last time at which a subject was seen, and still had not experienced the event. This is a special form of missing data, as we know that for individual $i$, the survival time exceeds $T_i$.

The censoring mechanism can be categorised as censored completely at random (CCAR), where censoring is completely independent of the survival mechanism; censoring at random (CAR), where the censoring is independent of the survival time, \emph{conditional} on the covariates that appear in the substantive model, for example treatment. If even after conditioning on covariates, censoring is dependent on the survival time, we say the censoring is not at random (CNAR).  CCAR and CAR are usually referred to as ignorable or non-informative censoring. 

Under non-informative censoring, we distinguish between two situations. The first one is when the outcome is fully observed, but we have missing values in the covariates that appear in the substantive model.  The second is when the survival time has missing values. 

In some settings covariates $X$ included in a Cox proportional hazards model have missing values, \citet{White2009}  
showed that when imputing either a normally distributed or binary covariate  $X$,  we should use a linear  (or logistic) regression imputation model, with $Y$ and the baseline cumulative hazard function as covariates. 
To implement this, we have to estimate the baseline cumulative hazard function. \citet{White2009}  suggested  that when covariate effects are small, baseline cumulative hazard  can be approximated by the Nelson-Aalen (marginal) cumulative hazard estimator, which ignores covariates and thus can be estimated using all subjects. Otherwise, the baseline cumulative hazard  function can be estimated within the FCS algorithm by fitting the Cox proportional hazards model to the current imputed dataset, and once we have this, we can proceed to impute $X$.

If the survival times are missing, according to  CAR, that is conditionally on the covariates used in our analysis model, then the results will be valid. MI can be helpful in situations where although the data are CAR, we are interested in estimating either marginal survival distributions, or our analysis model of interest includes fewer variables than required for the CAR assumption to be plausible. A parametric imputation model, for example the Weibull, log-logistic or lognormal is then used to impute the missing survival times, see \citet{ck13}.  

\vskip 2ex
\subsubsection{Clustered data}\label{Sec:clus} 
Clustered data may arise when the health economics study uses data from a multicentre RCT a cluster randomised trial, or indeed an observational study where observations are drawn from high level units such as hospitals or schools. In these settings but also those with panel or longitudinal data, it is important to recognise the resultant dependencies within the data. Methods have been developed for handling clustering in health economics studies including the use of multilevel (random effects) models \citep{Grieve2010, Gomes2011}, but these have not generally been used in settings with non-compliance. More generally there are methods to obtain CACE that accommodate clustering. These range from simply using robust SE estimation in an IV analysis \citep{Baiocchi2014}, to using multilevel mixture models within the principal stratification framework \citep{Frangakis2002, Jo2008}.  For extensions of principal stratification approaches to handle non-compliance at individual and cluster-level, see  \citet{Schochet2011}.

Regarding missing data, the FCS approach is not well-suited to proper multilevel MI and so, when using MI,  a joint modelling algorithm is used \citep{schafer97}. The R package \texttt{jomo}  can be used to multiply impute binary as well as continuous variables.  \citet{DiazOrdaz2014, DiazOrdaz2016} illustrate multilevel MI for bivariate outcomes such as costs and QALYs, to obtain valid inferences for cost-effectiveness metrics, and demonstrate the consequences of ignoring clustering in the imputation models.  

\section{Discussion}\label{Sec:discussion}

This chapter illustrates and critiques methods for handling non-compliance and missing data in health economics studies. Relevant methods have been developed for handling non-compliance in RCTs with bivariate \citep{DiazOrdaz2017}, or time to event endpoints \citep{Latimer2014a}. Methods for addressing missing data under MAR have been exemplified in health economics \citep{Briggs2003}, including settings with clustered data \citep{DiazOrdaz2014}.

Future studies are required that adapt the methods presented to the range of settings typically seen in applied health economics studies, in particular to settings with binary and time to event (duration) outcomes, and within longitudinal or panel datasets where data are missing at intermittent time points. To improve the way that non-compliance and missing data are handled, researchers are required to define the estimand of interest, transparently state the underlying assumptions, and undertake sensitivity analyses to departures from the missing data, and identification assumptions.

Future health economics studies will have increased access to large-scale linked observational data, which will include measures of non-adherence. In such observational settings, there will be new possible sources of exogenous variation such as genetic variants, resource constraints, or measures of physician or patient preference that offer possible instrumental variables for treatment receipt \citep{vonHinke2016, Brookhart2007}. Future studies will be required to carefully justify the requisite identification assumptions, but also develop sensitivity analyses to violation of these assumptions. Here, the health economics context is likely to provoke requirements for additional methodological development. For example, sensitivity analyses for the exclusion restriction \citep{Jo2002a, Jo2002b, Conley2012, Hirano2000}, may require refinement for settings where the exclusion restriction is satisfied for one endpoint (e.g. QALYs) but not for another (e.g. costs). Similarly, sensitivity analyses to the monotonicity assumption have been developed in the wider methodological literature \citep{Baiocchi2014, Small2017}, but they warrant careful consideration in the health economics context.

Further methodological research is required to allow for the more complex forms of non-adherence that may be seen in applied health economic studies. Compliance may not always be all-or-nothing, or time invariant. For example, there may be interest in the causal effect of the dose received. Here, available methods may be highly dependent on parametric assumptions, such as the relationship between the level of compliance and the outcome; alternatively  a further instrument is required in addition to  randomisation \citep{Dunn2007, Emsley2010}. 

A promising avenue of future research for providing less model-dependent IV estimates is to consider doubly robust estimators, such as targeted minimum loss estimation (TMLE) \citep{van2012targeted}, paired with ensemble machine-learning approaches, for example the so-called Super Learner \citep{van2007super}. TMLE approaches for IV models have recently been proposed \citep{Toth2016}.  In the health economics settings, TMLE has successfully been used for estimating the effects of continuous treatments \citep{Kreif2015}.

\section*{Acknowledgements}
We thank Mark Sculpher, Rita Faria, David Epstein, Craig Ramsey and the REFLUX study team for access to the data.\\
Karla DiazOrdaz was supported by UK  Medical Research Council Career development award in Biostatistics MR/L011964/1. This report is independent research supported by the National Institute for Health Research (Senior Research Fellowship, Richard Grieve, SRF-2013-06-016). The views expressed in this publication are those of the author(s) and not necessarily those of the NHS, the National Institute for Health Research or the Department of Health.

\clearpage
\begin{small}

\end{small}
\end{document}